\ifdefined\XeTeXversion\else\pdfoutput=1\fi
\documentclass[11pt]{article}

\usepackage[margin=1.15in]{geometry}
\usepackage{amsmath}
\usepackage{mathptmx}
\usepackage{graphicx}
\usepackage{booktabs}
\usepackage{microtype}
\usepackage[hidelinks]{hyperref}
\usepackage{url}
\makeatletter
\g@addto@macro\UrlBreaks{\do\a\do\b\do\c\do\d\do\e\do\f\do\g\do\h\do\i\do\j%
\do\k\do\l\do\m\do\n\do\o\do\p\do\q\do\r\do\s\do\t\do\u\do\v\do\w\do\x\do\y%
\do\z\do\0\do\1\do\2\do\3\do\4\do\5\do\6\do\7\do\8\do\9}
\makeatother
\newenvironment{note}{%
  \begin{list}{}{\setlength{\leftmargin}{1.5em}\setlength{\rightmargin}{1.5em}%
    \setlength{\topsep}{\medskipamount}\setlength{\parsep}{0pt}}%
  \item\small\textbf{Note.}\enspace}%
  {\end{list}}

\title{The Economics of AI Decoding Chips: Rebalancing Compute, Capacity,
and Bandwidth for Efficient LLM Inference}

\author{Michael J.\ Yuan (ByteFuture Inc.) \and Ju Long (Texas State University)}

\date{July 2026}

\begin{document}
\maketitle

\begin{abstract}
Every mainstream GPU is built compute-heavy and capacity-light: it pairs
enormous arithmetic throughput with too little memory to hold a modern model.
In contrast, large language model decoding requires little compute and a
large amount of memory: a GPU's floating-point units run at
single-digit-percent utilization during decoding, and the memory the workload
does need is sold only bundled with yet more compute. The
compute is recovered only at hyperscale, where Mixture-of-Experts (MoE) models
are spread across 96--320-GPU expert-parallel clusters serving thousands of
concurrent users, a scale available to a handful of operators. We formalize
the inefficiency with two fixed per-chip constants. F/B, the roofline ridge
point, determines whether the compute can be utilized; F/S, the compute
bundled with each GB of memory, determines how much compute must be bought. We
then argue for a rebalanced decode accelerator: less compute, far more
commodity memory, and a deliberately lower and cheaper bandwidth. The
Skymizer HTX-301, a purpose-built 28\,nm PCIe accelerator using commodity
DDR5, occupies that design point. Its entry cost is low. A single eight-chip
card holds DeepSeek-R1 671B for about \$19{,}000, and a 4U server of four
four-chip cards serves two users at a deterministic 20.3 tokens per second
each for about \$28{,}000. Either costs less than a single H100, while the
minimum GPU deployment for the model is an eight-GPU node near \$350{,}000.
Concurrency then scales out by adding hardware: eight 4U servers carry sixteen
users for about \$224{,}000, two-thirds of the node's price, with the cost per
token unchanged at about \$12 per million against the node's \$21. The
HTX-301's decisive advantage is a supply chain free of every rationed input:
it uses no high-bandwidth memory, no CoWoS, and no leading-edge logic.
\end{abstract}

\section{Introduction: the decode inefficiency}

Transformer inference has two phases with opposite hardware demands. In the
\emph{prefill} phase, the model processes the whole prompt in parallel, which
is compute-bound and uses a GPU's arithmetic units well. In the \emph{decode}
phase, the model generates output tokens one at a time; each step reads the
model's weights from memory and performs only about two floating-point
operations per weight byte read. Decoding is therefore memory-bound on all
practical hardware, and it is where interactive serving spends most of its
time and cost.

The mismatch this creates is structural, not incidental, and two ratios of a
chip's fixed resources govern it. The first is compute relative to memory
bandwidth, F/B, the roofline ridge point~\cite{roofline}. When F/B is low, the modest
arithmetic intensity of decoding saturates the arithmetic units at low
concurrency, the compute is fully utilized, and nothing is wasted. A
datacenter GPU, however, is engineered for the compute-bound regime and
carries a very high F/B, so during the decode phase most of its arithmetic units have nothing to do. The chip is limited by how fast it can stream weights, and the floating-point units idle at single-digit-percent utilization. The second ratio, compute relative to memory capacity, F/S,
matters even where F/B is acceptable, because the required capacity is set by
the model rather than chosen by the buyer. A high F/S converts a large
model's memory footprint directly into a mandatory purchase of compute:
holding a frontier Mixture-of-Experts model requires several GPUs for
capacity alone, which places datacenter-class compute in front of what may be
a single user. The memory itself, moreover, can be bought only bundled with
high-bandwidth memory (HBM) and CoWoS packaging that are themselves supply-constrained. The imbalance persists even in the newest products aimed at the edge. NVIDIA's RTX Spark, announced at Computex in June 2026 as the company's flagship AI platform for laptops and compact desktops, pairs 128\,GB of unified memory with roughly 300\,GB/s of bandwidth~\cite{rtxspark,rtxsparktoms}. As we show in \S\ref{sec:constants}, it therefore carries a ridge point even higher than the datacenter GPUs', one that no laptop workload can ever saturate.

This paper makes three moves. First (\S\ref{sec:constants}--\ref{sec:moe}), we
formalize the inefficiency with two fixed hardware constants and show that
Mixture-of-Experts models, now dominant at the frontier, make it far worse:
efficiency is recovered only at a concurrency, and a cluster scale, that only
hyperscalers can reach (\S\ref{sec:scale}). Second (\S\ref{sec:rebalance}), we
argue that the mass market below hyperscale needs a differently balanced chip,
one with less compute, more memory capacity, and a deliberately lower bandwidth
that the HBM/CoWoS supply crunch makes economically prudent. Third
(\S\ref{sec:instances}), we show that memory-rich
consumer devices approach this design point but do not reach it: their GPUs
carry more compute than decoding needs, pushing them just outside the corner,
and they lack the capacity to hold a frontier model. The purpose-built Skymizer HTX-301 fills both by pooling, serving DeepSeek-R1 671B on a \$28{,}000 4U system, against the \$350{,}000 eight-GPU node the model otherwise requires.

\section{Two hardware constants: F/B and F/S}
\label{sec:constants}

Let $F$ be a chip's peak compute throughput, $S$ its memory capacity, and $B$
its memory bandwidth. Two ratios, fixed at design time, govern decoding
efficiency, and they answer different questions. F/B is genuinely a question
of proportion: it determines what fraction of the compute a workload can use.
F/S, because the required capacity is set by the model rather than chosen by
the buyer, is in effect a question of absolute quantity: it determines how
much compute must be bought at all. Throughout, F/B uses each device's peak
dense 8-bit tensor throughput (FP8/INT8) over its peak memory bandwidth, the
pairing relevant to quantized decoding with 8-bit activations, and F/S uses
peak dense BF16 tensor throughput per GB. A datasheet throughput quoted ``with
sparsity,'' the 2:4 structured-sparsity mode that deployed LLMs do not use, is
halved to the dense rate.

\paragraph{F/B: the ridge point, and the concurrency it demands.}
The roofline model of Williams, Waterman, and Patterson states that a workload
uses a chip's compute only when its arithmetic intensity (FLOP per byte of
memory traffic) exceeds the chip's \emph{ridge point},
$F/B$~\cite{roofline}. Single-user decoding has an arithmetic intensity of only a few FLOP/byte. Batching $K$ concurrent users multiplies it by roughly $K$, because they share each weight read. Measurement studies confirm that decoding on GPUs remains memory-bound even at large batch sizes~\cite{recasens}. So the concurrency required to
saturate a chip is approximately its ridge point divided by the per-user
intensity. Datacenter GPUs have very high ridge points (H100 $\approx 591$,
B200 $\approx 584$), so for INT4 decoding they need on the order of 140--150
concurrent users before the compute is used at all. The pattern is not
confined to GPU vendors: Google's TPU~v7, with 4,614 FP8 TFLOPs over
7.37\,TB/s of HBM, sits at $\approx 626$~\cite{tpuv7}. The two Spark platforms
are instructive: they are memory-rich yet even more bandwidth-starved
relative to their compute (F/B $\approx 833$--916, the
highest ridge points in this comparison), so they inherit a concurrency
requirement beyond even the datacenter GPUs'. The RTX Spark, announced at
Computex in June 2026 and built on the same N1 silicon as the DGX Spark,
brings exactly this profile to consumer laptops and compact desktops, pairing
128\,GB of unified LPDDR5X with a reported $\sim$300\,GB/s of
bandwidth~\cite{rtxspark,rtxsparktoms}. Among the DRAM-based devices, none
keeps bandwidth in proportion to compute: the lowest ridge point is the
H200's $\approx 412$, which still requires roughly 103 concurrent users. Two
SRAM-based accelerators, the Groq LPU and the Cerebras WSE-3, do achieve
genuinely low ridge points (F/B $\approx 6$--9) by keeping all memory
on-die~\cite{groqcard,wse3,cerebrasinf}, which is why they excel at
single-user latency; their difficulty lies on the other axis.

\paragraph{F/S: the compute you are forced to buy per GB.}
Even a chip whose F/B were acceptable can still be the wrong purchase, because
the second ratio fixes the size of the purchase itself. Decoding requires
bytes to be held and bytes to be streamed, not floating-point operations, yet
a chip's F/S (TFLOPS per GB of capacity) dictates how much compute is bundled
with each GB of memory. Since the capacity a deployment must provision equals
the model's footprint, the total compute acquired is the product of F/S and
the model size: a high F/S converts a large model directly into a large, and
for decoding largely useless, absolute quantity of compute. Datacenter GPUs,
workstation GPUs, and the TPU~v7 sit at F/S $\approx 5$--14 TFLOPS/GB
(Table~\ref{tab:constants}). Adding commodity memory helps at the margin.
NVIDIA's DGX Spark and its consumer sibling, the RTX Spark, reach $\approx 1$, an order of magnitude below the datacenter GPUs. That is the lowest F/S in this comparison, and every DRAM-based device here still carries far more compute than decoding can use. The SRAM-based accelerators occupy
the opposite extreme. The Groq LPU carries 230\,MB of SRAM per chip and the
Cerebras WSE-3 carries 44\,GB per wafer, so their F/S is roughly 817 and
2,841 TFLOPS/GB, two to three orders of magnitude above the
GPUs'~\cite{groqcard,wse3}. Holding DeepSeek-R1 at 4-bit precision, roughly
400\,GB, would therefore require on the order of 1,700 Groq chips or nine
wafer-scale systems for capacity alone. A low ridge point does not rescue a
design whose per-device capacity is so small that the mandatory compute
purchase becomes enormous; these architectures buy minimum latency at
maximum hardware, the mirror image of the GPU's failure rather than a
correction of it.

\begin{table}[t]
\centering
\caption{The table reports, for representative accelerators, memory capacity,
the two fixed decode constants F/B (ridge point, FLOP/byte) and F/S
(TFLOPS/GB), and the derived concurrency to saturate compute at INT4 decode
(users to saturate). The two SRAM-based devices below the rule keep all memory
on-die, so their bandwidth is aggregate on-die SRAM bandwidth, not directly
comparable to DRAM. Specifications are from vendor
datasheets~\cite{nvb200,mi300x,tpuv7,dgxspark,rtxspark,groqcard,wse3,cerebrasinf};
the RTX Spark's DRAM bandwidth ($\sim$300\,GB/s) is press-reported~\cite{rtxsparktoms},
and the TPU v7's F/S takes BF16 as half its published FP8 rate.}
\label{tab:constants}
\begin{tabular}{lrrrr}
\toprule
Device & Memory (GB) & F/B (ridge) & Users to saturate & F/S (TFLOPS/GB) \\
\midrule
H100          & 80  & 591 & 148 & 12.4 \\
B200          & 180 & 584 & 146 & 12.5 \\
H200          & 141 & 412 & 103 & 7.0  \\
MI300X        & 192 & 493 & 123 & 6.8  \\
TPU v7 (Ironwood) & 192 & 626 & 157 & 12.0 \\
RTX 5090      & 32  & 468 & 117 & 13.1 \\
RTX PRO 6000  & 96  & 558 & 140 & 5.2  \\
DGX Spark     & 128 & 916 & 229 & 0.98 \\
RTX Spark (N1X) & 128 & 833 & 208 & 0.98 \\
\midrule
Groq LPU      & 0.23 & 9  & 2   & 817  \\
Cerebras WSE-3 & 44  & 6  & 1   & 2{,}841 \\
\bottomrule
\end{tabular}
\end{table}

\begin{figure}[t]
\centering
\includegraphics[width=\linewidth]{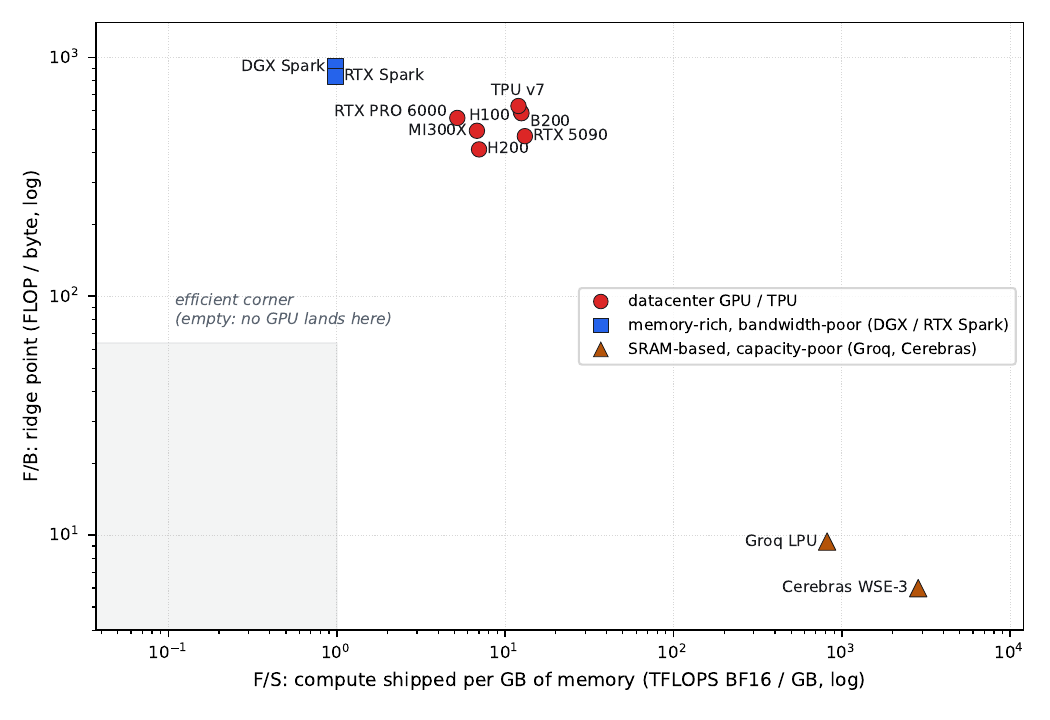}
\caption{The figure plots each accelerator by F/S (horizontal, TFLOPS/GB) and
F/B (vertical, FLOP/byte) on log axes, with marker shape encoding device class
(see legend). The shaded region is the efficient corner, low on both axes; no
device here lands in it.}
\label{fig:corner}
\end{figure}

Figure~\ref{fig:corner} plots the two constants together, and four classes
emerge. The datacenter and workstation GPUs cluster with the TPU in the
high-F/B, high-F/S upper right: abundant compute, scarce capacity, and a
concurrency requirement that idles the compute below hyperscale. That the TPU~v7 lands in the same cluster is informative rather than coincidental. The TPU's real differentiation lies in its pod-scale optical interconnect, its performance per watt, and its owner's cost structure~\cite{tpuv7}. These are axes the two constants do not measure, and all of them serve hyperscale deployment. Any design that pairs HBM with maximal leading-edge
compute arrives at this corner regardless of vendor. The two memory-rich
devices, the DGX and RTX Spark, carry the highest ridge points of all, and
the SRAM machines occupy the mirror-image lower right, low F/B at extreme
F/S. The efficient corner, low on both axes, is empty: not one of these
accelerators lands there. The corner marks the decode-matched region: an F/S
below about 1\,TFLOP/GB (memory-rich) and a ridge point below about 64. At the
$\approx$4 FLOP/byte intensity of INT4 decode, a ridge of 64 saturates at
about sixteen concurrent users ($64/4$), the on-prem concurrency this paper
targets; a higher ridge needs more users than that market provides. What kind of chip \emph{would} fill that corner,
and what already does, is the subject of
\S\ref{sec:rebalance}--\ref{sec:instances}.

\section{The MoE tax and why efficiency requires scale}
\label{sec:moe}

\begin{figure}[t]
\centering
\includegraphics[width=0.92\linewidth]{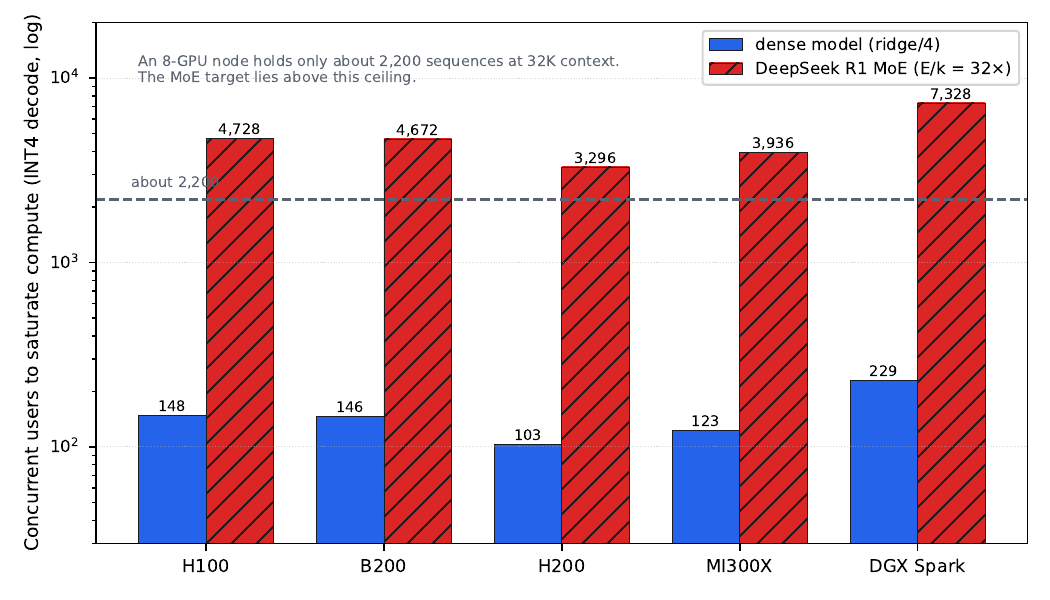}
\caption{The figure plots, for each GPU, the concurrency needed to saturate
its compute on a dense model (left bars) versus DeepSeek-R1 (right bars,
hatched), at INT4 decode on a log scale. The dashed line marks a single 8-GPU
node's KV-cache batch ceiling.}
\label{fig:moetax}
\end{figure}

Frontier open models are now overwhelmingly Mixture-of-Experts, and MoE
sharply worsens the concurrency problem. An MoE layer routes each token to
$k$ of $E$ experts, so a batch of $K$ users spreads across the experts and
each expert sees only $K \cdot k/E$ tokens. The effective arithmetic intensity
is therefore divided by the sparsity ratio $E/k$, which we call the ``MoE
tax.'' For DeepSeek-R1 (256 experts, top-8) the tax is
$32\times$~\cite{deepseekv3}. The concurrency needed to saturate a GPU is
multiplied by the same factor: an H100 that needs $\approx 148$ dense users
would need $\approx 4{,}700$ concurrent R1 users (Figure~\ref{fig:moetax}).
This target exceeds what the KV-cache memory of an 8-GPU node permits, which
is on the order of 2{,}200 concurrent sequences at 32K context. At any
feasible batch, moreover, each expert receives fewer tokens than the
$\sim$128-token tile that saturates the grouped-GEMM
kernels~\cite{deepgemm}. A single node running a large MoE model therefore
\emph{cannot} reach the concurrency that would use its compute, and it stays
memory-bound at very low utilization. DeepSeek-R1 on an 8-GPU node measures
roughly 0.6\% model-FLOPS utilization (MFU, the fraction of the device's peak
floating-point throughput that the workload actually uses), falling to about 0.03\% at a single user~\cite{dzhsurf,tensoreconomics}. MFU follows directly from
measured throughput:
\begin{equation}
\mathrm{MFU} \;=\; \frac{2\,P_{\mathrm{active}}\,T}{F_{\mathrm{peak}}},
\end{equation}
where $T$ is the aggregate decode throughput in tokens per second,
$P_{\mathrm{active}}$ is the number of active parameters per token (37B for
DeepSeek-R1), the factor of two counts one multiply and one accumulate per
parameter, and $F_{\mathrm{peak}}$ is the system's aggregate peak dense BF16
throughput. The single-node figure follows: 620 tokens per second across
eight H100s gives $\mathrm{MFU} = 2 \times 37{\times}10^{9} \times 620 \,/\,
(8 \times 989.5{\times}10^{12}) \approx 0.6\%$.

Efficiency is recovered only by \emph{expert parallelism} (EP): spreading the
experts across many GPUs so that each holds few experts and the aggregated
traffic from thousands of users fills every expert. DeepSeek's own production
decode unit places its experts across 144 H800s (EP144), roughly two routed
experts per GPU, which lifts arithmetic intensity by an order of magnitude and
recovers MFU to about 20--25\%~\cite{deepseekday6,lmsys}. Independent reproductions
reach similar figures on 96--320-GPU clusters~\cite{lmsys}. The physics thus
forces a stark choice for large MoE serving: either operate at hyperscale on a
96--320-GPU EP cluster, or accept single-digit-percent utilization.

Viewed through the two constants, the MoE tax is the F/S problem in its
sharpest form. Capacity must be provisioned for every expert, so the compute acquired scales with total parameters, yet each token exercises only the active fraction. A DeepSeek-R1 deployment buys compute in proportion to 671B parameters, while each user's decode stream works 37B of them per token~\cite{deepseekv3}. The only way to use the purchased compute is to
raise arithmetic intensity toward the ridge point, and the GPU's expensive
high bandwidth makes that possible only by aggregating thousands of
concurrent users. A chip whose cost rests on high F and high B is therefore
structurally unsuited to low-concurrency serving: the compute idles, and the
bandwidth that was paid for to feed it serves only a handful of streams.

\section{The cost of scale, and the market gap}
\label{sec:scale}

\begin{figure}[t]
\centering
\includegraphics[width=0.92\linewidth]{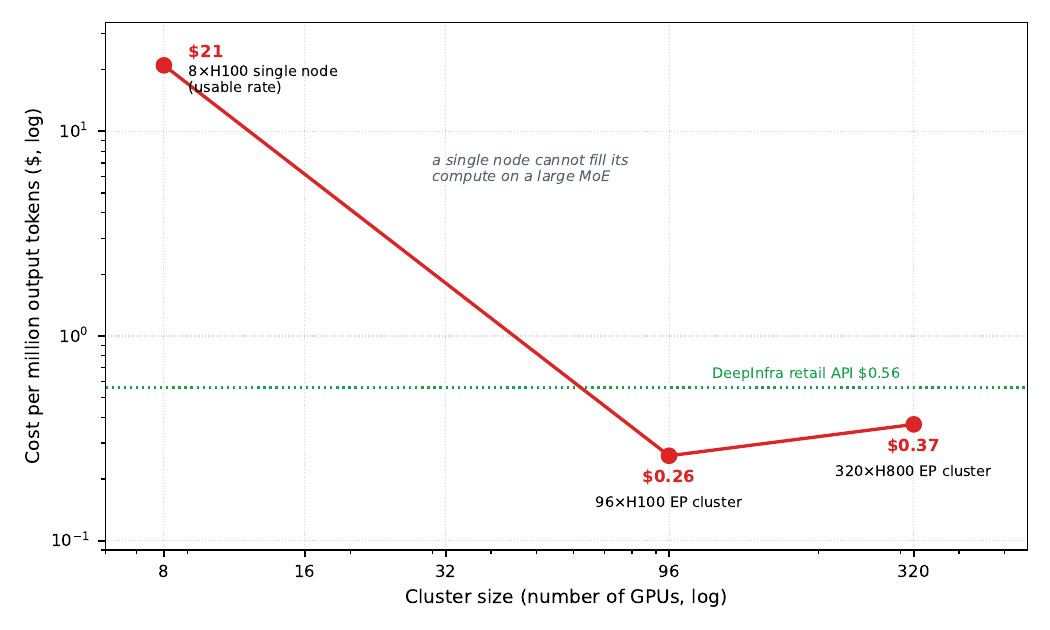}
\caption{The figure plots cost per million output tokens for DeepSeek-R1
against GPU cluster size (log-log). The three points are production
deployments (8, 96, and 320 GPUs); the dotted line marks a retail API price.}
\label{fig:costcurve}
\end{figure}

Scale resolves the utilization problem but concentrates the market.
Figure~\ref{fig:costcurve} plots the cost per output token for DeepSeek-R1
against cluster size. Here and throughout the paper, cost per million output
tokens is computed with a single ownership model:
\begin{equation}
\label{eq:costmodel}
\text{\$/M} \;=\; \frac{C_{\mathrm{hour}}}{R},
\qquad
C_{\mathrm{hour}} \;=\; \frac{P_{\mathrm{acq}}}{2 \times 8{,}760\ \mathrm{h}}
\;+\; W \cdot c_{e},
\end{equation}
where $R$ is the system's output volume in millions of tokens per hour,
$C_{\mathrm{hour}}$ is its hourly cost, $P_{\mathrm{acq}}$ is its acquisition
price amortized over two years of continuous operation (17,520 hours), $W$ is
its sustained power draw in kW, and $c_{e} = \$0.20$ per kWh is the price of
electricity. The power draw is taken well below each node's maximum rating,
since memory-bound decoding does not exercise the arithmetic units; a longer
amortization lowers every hourly cost, and every cost per token, roughly
proportionally, without changing the comparisons. Under this model, an 8-GPU
H100 node at
$P_{\mathrm{acq}} = \$350\mathrm{K}$ and $W = 5.0$\,kW costs
$C_{\mathrm{hour}} = \$21.0$, or about \$2.6 per GPU-hour, in line with
prevailing rental rates. A single node cannot fill its compute. At a usable interactive rate it serves
about sixteen users at 17 tokens per second each, some 270 tokens per second in
aggregate, so $R = 0.98$ and Equation~\eqref{eq:costmodel} gives about \$21 per
million tokens. It can be driven cheaper, to \$9.41, but only by batching to a
hundred users at 6 tokens per second each, below reading speed and not a rate
anyone would serve on~\cite{dzhsurf}. A 96-GPU EP cluster decodes 2,787 tokens
per second per GPU, so $R = 963$, and its twelve nodes cost \$252 per hour,
which gives \$0.26 per million tokens~\cite{lmsys}. The improvement from a
usable single node to the cluster is roughly $80\times$, and it comes almost
entirely from finally using the compute.

But an EP cluster is a
\$4--13M capital commitment that must be kept saturated with a continuous
stream of thousands of concurrent users to realize that cost. Only operators
with hyperscale demand can amortize it.

This leaves a large, fast-growing, and supply-constrained market underserved.
Sovereign clouds, enterprise on-premises deployments, telco edge sites, and
single-tenant appliances have real demand for frontier MoE inference but
nowhere near hyperscale concurrency. Both of their current options are poor:
they can buy an 8-GPU node and run it at under 1\% MFU, paying for compute
that idles, or they can rent from a provider at a large markup. The engineering points to a single opening:
a chip whose cost is governed by memory rather than compute, and which reaches
efficiency at low concurrency.

\section{A rebalanced design: less F, more S, deliberately less B}
\label{sec:rebalance}

The remedy follows directly from the two constants. To serve the decode phase
efficiently outside hyperscale, the silicon must be rebalanced along three
axes.

\begin{itemize}
\item \textbf{Less compute (F).} Decoding at realistic edge concurrency cannot
exercise a datacenter GPU's arithmetic. Cutting F removes stranded units and
lowers both F/S and the ridge point F/B, and because compute area dominates
cost on advanced nodes, it removes the main reason the chip must be built on
an expensive process. It is the cut in F, not the lower B, that keeps the
ridge point low.
\item \textbf{More memory capacity (S), from commodity DRAM.} The binding
requirement is holding the whole model, including every expert of an MoE.
Provisioning large capacity from commodity DRAM, rather than HBM, collapses
F/S and sidesteps the HBM allocation queue. Capacity can be configured per
card, as in a personal computer, so one design serves models from a few
billion to a trillion parameters.
\item \textbf{Deliberately lower bandwidth (B), as a considered trade.} Lower
$B$ does not improve the ridge point; for a given F, reducing B raises F/B.
The benefit of lower B lies in cost and supply. It allows commodity DRAM in
place of HBM, whose supply is the tightest constraint in the AI hardware
chain. Because HBM and CoWoS packaging are sold out into
2026--27~\cite{epochsupply,semianalysis}, a design that needs neither can
actually be built and shipped at volume. The cost of lower B is slower
single-user token generation, and this cost sets a floor: memory-bound
decoding delivers tokens in proportion to B, so B must stay high enough for
an acceptable interactive rate. The bandwidth is deliberate in exactly this
sense. It is chosen as low as the token-rate requirement allows, but no
lower, and the remaining per-user slowness is recovered in aggregate by
stacking cheap cards into a small decode pool.
\end{itemize}

The design point is thus a slow but efficient, memory-heavy, commodity-supply
accelerator. It concedes the hyperscale market and single-user
latency-critical work, and in exchange it decodes frontier models cheaply for
everyone below hyperscale, on a supply chain that is not rationed. This corner
is empty of GPUs (Figure~\ref{fig:corner}), and of shipping products too. The
next section examines the two that come nearest: a consumer workstation just
outside the corner, and a purpose-built accelerator that lands squarely in it.

\section{Filling the corner}
\label{sec:instances}

No datacenter GPU reaches the efficient corner of Figure~\ref{fig:corner}
(\S\ref{sec:constants}): each is built compute-heavy, so at the low concurrency
of on-premises serving most of its arithmetic sits idle. A machine built for a
single user ought to do better, since it need not batch a crowd to earn its
keep. Of the shipping consumer devices, the closest to the corner is Apple's
Mac Studio (M3 Ultra), a popular and representative platform for personal LLM
inference. It
pairs unified LPDDR5 at about 819\,GB/s with an 80-core GPU of roughly
65\,BF16 TFLOPS, twice its 32.8\,TFLOPS FP32 rate~\cite{m3ultragpu}. That gives
it an F/S near 0.68 and a ridge point near 80, both far below every datacenter
GPU. Yet even it falls short. The ridge still sits just above the corner,
because even a machine sold for one user carries more compute than decoding can
exercise. The AI PCs aimed at the same market, NVIDIA's DGX and RTX Spark,
hold more memory than the Mac but pair it with far less bandwidth, so they sit
well outside the corner (Figure~\ref{fig:corner}). And capacity closes the door regardless.
Holding DeepSeek-R1 671B at 4-bit needs about 400\,GB. A single Mac Studio now
tops out at 96\,GB, after Apple withdrew the 128, 256, and 512\,GB options
during the 2026 memory shortage, and macOS cannot pool memory across
units~\cite{macstudio}. It cannot hold the model.

The Mac Studio is nonetheless a sharp illustration of the F/S problem, and a capable device for the models that do fit its memory. Its 96\,GB arrives bundled with a leading-edge 3\,nm compute die that decoding cannot exercise, so a buyer who wants more memory must pay for more of that stranded compute. It serves models up to a few hundred billion parameters well; only the frontier exceeds it.

\begin{figure}[t]
\centering
\includegraphics[width=\linewidth]{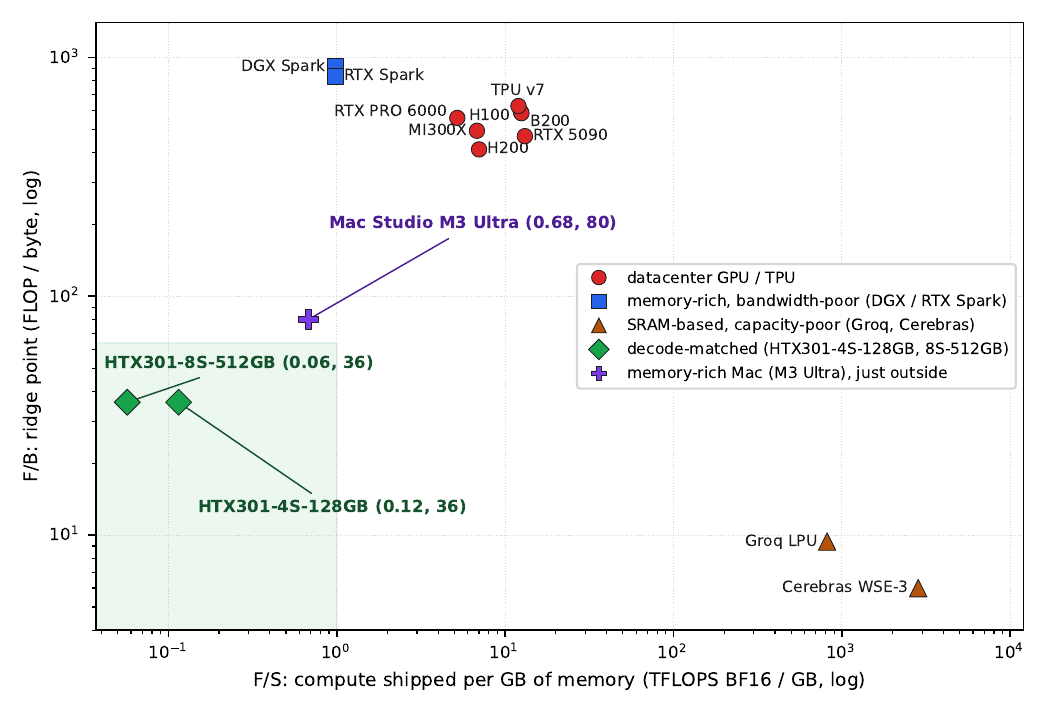}
\caption{The figure repeats the F/S versus F/B map of Figure~\ref{fig:corner}
with both HTX-301 configurations (HTX301-4S-128GB and HTX301-8S-512GB) inside
the corner and the Mac Studio just outside it. Marker shape encodes device class (see legend). The map plots
the two efficiency ratios only, not capacity.}
\label{fig:corner_filled}
\end{figure}

What fills both is a device built to pool. The Skymizer HTX-301 is a
purpose-built 28\,nm PCIe accelerator using commodity DDR5, and it comes
in two cards~\cite{htx301}. The HTX301-4S-128GB carries four inference chips and
128\,GB; the HTX301-8S-512GB carries eight chips and 512\,GB. Each chip supplies
3.68\,TFLOPS of BF16 compute over a 102\,GB/s channel. The four-chip card thus
delivers about 15\,TFLOPS across 410\,GB/s at 164\,W, the eight-chip card about
29\,TFLOPS across 819\,GB/s at 328\,W. Both sit well inside the efficient corner
(Figure~\ref{fig:corner_filled}): the four-chip card at F/B $\approx 36$ and
F/S $\approx 0.12$, the eight-chip card at the same ridge point but F/S
$\approx 0.06$, deeper still. Both sit an order of magnitude below the GPUs on
the ridge point and two orders below on compute-per-GB, and need no HBM or
CoWoS.

Each card's cost is chips plus commodity memory. Four chips at about \$1{,}000
each and 128\,GB of DDR5 at about \$20 per GB make the four-chip card roughly
\$6{,}600 in parts, near \$7{,}000 built. Eight chips and 512\,GB make the
eight-chip card roughly \$18{,}000 in parts, near \$19{,}000 built. Pooling four
of the four-chip cards in a 4U server gives 512\,GB for about \$28{,}000, enough
to hold DeepSeek-R1 671B at 4-bit. It is rated at a deterministic 20.3
tokens per second for each of two concurrent users.

\begin{note}
The ridge points for the HTX-301 and the Mac Studio use their BF16 rate,
because neither publishes an accelerated 8-bit tensor rate, while the GPU
ridge points use 8-bit. The choice of basis shifts the absolute numbers but
not the ordering: the HTX-301 stays roughly an order of magnitude below every
GPU and well inside the corner, and the Mac Studio stays between the two,
just outside it.
\end{note}

\subsection{What it takes to run DeepSeek-R1 671B}

To serve the same model on GPUs, an operator cannot start small. A single
H100 carries 80\,GB, far short of the model's $\sim$400\,GB, so the minimum viable GPU deployment is a full eight-GPU node~\cite{dzhsurf} with 640\,GB of HBM, about \$350{,}000. The HTX-301 has no such floor: even its smallest deployment, a single card, runs the model for less than the price of one H100 GPU. The GPU
path requires eight of them before it runs at all
(Table~\ref{tab:run671}).

\begin{table}[t]
\centering
\caption{The table reports, for each deployment that serves DeepSeek-R1 671B
(Q4) in the decode phase, its concurrency (concurrent users), acquisition
cost, and cost per million output tokens at that concurrency. Prices are
system-integrated.}
\label{tab:run671}
\begin{tabular}{lrrr}
\toprule
System & Concurrency & Acq.\ cost & \$/M \\
\midrule
HTX301-8S-512GB & 1 & \$19K & \$20 \\
4U HTX301-4S-128GB & 2 & \$28K & \$12 \\
4U HTX301-4S-128GB $\times$8 & 16 & \$224K & \$12 \\
8$\times$H100 node & 16 & \$350K & \$21 \\
\bottomrule
\end{tabular}
\end{table}

\paragraph{The entry cost of running the model.}
The price of admission comes before the cost per token. Below
an eight-GPU node, the GPU path cannot serve DeepSeek-R1 671B at all. Every
HTX-301 deployment can, and for far less: about \$19{,}000 for a single eight-chip card, \$28{,}000 for a four-card 4U server, and \$224{,}000 for eight of those servers. The largest is still two-thirds the cost of the minimum GPU node; the smallest is about one-eighteenth of it. For a sovereign deployment, an on-premises appliance, or an edge site that needs the model for a handful of users, that gap is the difference between a capital project and a workstation purchase.

\paragraph{Matched throughput: cheaper per token, and deterministic.}
When more concurrency is needed, the HTX-301 scales by replicating 4U units. Eight of them, thirty-two cards, serve sixteen concurrent users at the same fixed 20.3 tokens per second, matching an eight-GPU node's concurrency at about \$224{,}000 against \$350{,}000, roughly two-thirds the capital. Because the pool scales linearly, its cost per token does not change with size. Under the ownership model of Equation~\eqref{eq:costmodel}, the HTX-301 decodes at about \$12 per million output tokens at any scale, against roughly \$21 for the GPU node at the same sixteen users (Table~\ref{tab:run671}), a little over half as much.

\begin{note}
That the gap is under twofold rather than an order of magnitude is worth
stating. The HTX-301's decode-matched efficiency is offset by its mature
28\,nm process, which packs far less compute into a given area and power
budget than the H100's leading-edge node. The per-token saving is therefore
real but modest, and the stronger case for the HTX-301 is its entry cost and
its supply chain.
\end{note}

The HTX-301 rate is also deterministic: every user sees the same 20.3 tokens
per second regardless of who else is on the box, whereas the GPU node's
per-user rate moves with batch composition.

A single eight-chip card is the third deployment. It holds DeepSeek-R1 671B for
about \$19{,}000 and serves one user. With half the 4U's chips and half its
users, each user has the same bandwidth behind it; what the single 8S user
lacks is a batch to share the read. Decoding a Mixture-of-Experts model streams
the active experts once per token, and the 4U's two users split that read where
the lone 8S user pays it whole. Its per-token read is R1's full 37 billion
active parameters, about 18.5\,GB at 4-bit, against roughly 14.5\,GB per user
when two share it. The rate falls in that ratio, from the 4U's 20.3 to about 16
tokens per second. By Equation~\eqref{eq:costmodel} that is about \$20 per
million (Table~\ref{tab:run671}), close to the GPU node's \$21 and well above
the 4U's \$12. A single card sets the low entry price; a single user sets the
high cost per token.

\begin{note}
A decode step reads two kinds of weight per token. The shared weights
(attention, the router, and the always-resident shared expert) are about
8\,GB at 4-bit and identical for every token. The routed experts a token
activates, its top-8 of 256, are about 10\,GB and differ from token to token.
Batching amortizes the first but not the second. Two users' tokens generated
in one step read the 8\,GB of shared weight once, so each token pays 4\,GB
rather than 8. The two tokens activate almost disjoint expert sets, though, so
the step fetches about sixteen distinct experts, roughly 20\,GB, for its two
tokens, about 10\,GB each. The per-token read therefore falls from 18.5\,GB
for a lone user to about 14.5\,GB at two users, not to the 9.25\,GB a dense
model would give, where every weight is shared. Batching a sparse model
recovers only the shared weights; the expert traffic rides at full price,
which is why decoding stays memory-bound where a dense model, sharing every
weight, would become compute-bound.
\end{note}

\paragraph{Commodity memory, and a commodity logic node.}
The HTX-301's cost and supply-chain advantages come from the same two
choices. It uses commodity DDR5, a fraction of HBM's per-GB price and free
of CoWoS packaging, and HBM and CoWoS are precisely the two constraints
throttling GPU supply~\cite{epochsupply,semianalysis}. It is built on a
mature 28\,nm process that costs roughly $6\times$ less per wafer than a
leading-edge node, needs no EUV~\cite{wafer}, and draws on abundant capacity
that nothing else is contending for. It therefore competes for none of the
rationed inputs, on either the memory or the logic side. The Mac Studio, by
contrast, is memory-rich but built on a leading-edge 3\,nm process. That
process competes for the same scarce advanced-node capacity as high-end
GPUs, so the Mac is not insulated from the AI buildout on the logic side. In
any case it cannot hold the model.

These advantages come with real trade-offs. The GPU node decodes faster per
user, runs a mature and battle-tested software stack, and ships with vendor
support. For an operator who already needs many concurrent users and can
amortize the node, or who serves at hyperscale, it is the sounder buy
(\S\ref{sec:scale}). The HTX-301 is slower per user, and its single-stream
ceiling is low by design. What it offers in exchange is a frontier model at the
price of a workstation, on a supply chain that competes for none of the
rationed inputs. For an operator whose binding constraint is capital,
server integration, supply resilience, or a deterministic latency guarantee,
that exchange is decisive. The use cases and limits below
(\S\ref{sec:usecases}) map it onto concrete deployments.

\section{Use cases and limits}
\label{sec:usecases}

The rebalanced accelerator is an edge, on-premises, and sovereign product, not
a datacenter GPU replacement. It fits three shapes of demand. As an
\emph{edge appliance}, it occupies the desktop-and-rack niche of the Mac
Studio and the DGX Spark, but as a poolable PCIe card that reaches
frontier-model capacity those sealed units cannot. In an \emph{on-premises or edge server}, it can
form the decode half of a disaggregated deployment, with a GPU handling the
compute-bound prefill phase and a pool of HTX-301 cards handling the
memory-bound decode phase~\cite{distserve,splitwise}. And
for \emph{sovereign or single-tenant} operators, it offers frontier-model
serving on a supply chain free of the HBM/CoWoS allocation queue.

It does not fit two regimes. Hyperscale serving of thousands of concurrent
users is already efficient on 96--320-GPU EP clusters at roughly 20\% MFU and
\$0.2--0.3 per million tokens~\cite{aar1}, which the HTX-301 does not beat. And
latency-critical single-user work needing more than about 20 tokens/second is
not served by a deliberately low-bandwidth part. A node shrink to 6\,nm would
raise single-user speed and widen the addressable market, but it re-enters the
EUV and LPDDR5X cost tier and so softens the cheap-and-abundant thesis; it
would be a performance-oriented successor rather than the cost disruptor. The
28\,nm part remains the supply-chain and cost proposition.

\section{Conclusion}

LLM decoding needs little compute and much memory, yet the mainstream GPU
carries far more compute than decoding can use, with the needed memory
obtainable only through scarce HBM and CoWoS. That mismatch is tolerable at
hyperscale, where 96--320-GPU expert-parallel clusters finally batch enough
users to put the compute to work, and expensive everywhere else. The two fixed
constants F/B and F/S locate the problem precisely, and the MoE tax shows why
frontier models make it acute below hyperscale. The remedy is to rebalance the silicon with less compute, more commodity memory, and a deliberately lower bandwidth, an economy that the supply crunch makes prudent. The result is a slow but cheap, memory-heavy accelerator that decodes
frontier models efficiently at low concurrency. Consumer memory-rich devices
approach this corner but do not reach it. Their GPUs carry more compute than
decoding needs, so the Mac Studio sits just outside it. Its 96\,GB ceiling
and lack of pooling keep it below frontier-MoE capacity in any case. The purpose-built
Skymizer HTX-301 reaches both the corner and the capacity by pooling four
28\,nm commodity-DDR5 cards in a 4U server. It serves DeepSeek-R1 671B for less than the price of a single H100, where the GPU path needs an eight-GPU node to run the model at all. Its deeper advantage is the supply chain: with
neither HBM nor CoWoS nor leading-edge logic, nothing it needs is rationed by
the AI buildout. It undercuts a GPU node by roughly an order of magnitude in
capital for the low-concurrency frontier-MoE workloads that dominate the edge
and on-premises market.

\section*{Acknowledgments}

The authors thank Luba Tang, Chief Executive Officer of Skymizer Inc., and
Panos Kao, Director of Product Management at Skymizer, for helping us
understand the HTX-301 cards' configurations, bill-of-materials prices, and
capabilities.

\end{document}